\documentclass{emulateapj}
\usepackage{amsmath}

\shorttitle{On the detectability of lightcurves of KBOs}
\shortauthors{Lacerda \& Luu}
\journalinfo{Icarus 161 (2003) 174--180}
\submitted{Feb 2003}
\begin{document}

\title{On the detectability of lightcurves of \mbox{Kuiper Belt
    Objects}}

\author{Pedro Lacerda}
\affil{Leiden Observatory, University of Leiden,
  Postbus 9513, NL-2300 RA Leiden, Netherlands}
\email{placerda@strw.leidenuniv.nl}
\and
\author{Jane Luu}
\affil{MIT Lincoln Laboratory, 244 Wood Street, Lexington, MA 02420, USA}
\email{luu@ll.mit.edu}

\begin{abstract}
  We present a statistical study of the detectability of lightcurves
  of Kuiper Belt objects (KBOs). Some Kuiper Belt objects display
  lightcurves that appear "flat", i.e., there are no significant
  brightness variations within the photometric uncertainties. Under
  the assumption that KBO lightcurves are mainly due to shape, the
  lack of brightness variations may be due to (1) the objects have
  very nearly spherical shapes, or (2) their rotation axes coincide
  with the line of sight.  We investigate the relative importance of
  these two effects and relate it to the observed fraction of ``flat''
  lightcurves.  This study suggests that the fraction of KBOs with
  detectable brightness variations may provide clues about the shape
  distribution of these objects. Although the current database of
  rotational properties of KBOs is still insufficient to draw any
  statistically meaningful conclusions, we expect that, with a larger
  dataset, this method will provide a useful test for candidate KBO
  shape distributions.
\end{abstract}



\section{Introduction}

    The Kuiper Belt holds a large population of small objects which
    are thought to be remnants of the protosolar nebula
    \citep{jewitt}.  The Belt is also the most likely origin of other
    outer solar system objects such as Pluto-Charon, Triton, and the
    short-period comets; its study should therefore provide clues to
    the understanding of the processes that shaped our solar
    system. More than 650 Kuiper Belt objects (KBOs) are known to date
    and a total of about $10^{5}$ objects larger than 50~km are
    thought to orbit the Sun beyond Neptune \citep{jewitt3}.

\begin{figure}
      \epsscale{1.0}
      \plotone{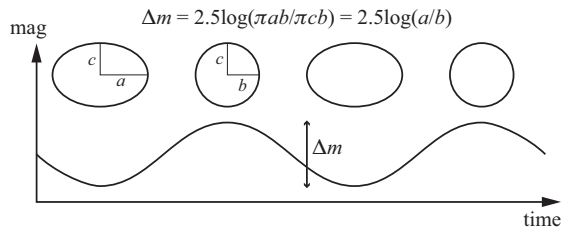}
      \caption{The lightcurve of an ellipsoidal KBO observed at aspect
	angle $\theta=\pi/2$. Cross-sections and lightcurve are
	represented for one full rotation of the KBO. The amplitude,
	$\Delta m$, of the lightcurve is determined for this
	particular case. See text for the general ex\-pression.
	\label{FigLC}}
    \end{figure}

    One of the most fundamental ways to study physical properties of
    KBOs is through their lightcurves. Lightcurves show periodic
    brightness variations due to rotation, since, as the KBO rotates
    in space, its cross-section as projected in the plane of the sky
    will vary due to its non-spherical shape, resulting in periodic
    brightness variations (see Fig.~\ref{FigLC}). A well-sampled
    lightcurve will thus yield the rotation period of the KBO, and the
    lightcurve amplitude has information on the KBO's shape. This
    technique is commonly used in planetary astronomy, and has been
    developed extensively for the purpose of determining the shapes,
    internal density structures, rotational states, and surface
    properties of atmosphereless bodies. These properties in turn
    provide clues to their formation and collisional environment.

    Although lightcurves studies have been carried out routinely for
    asteroids and planetary satellites, the number of KBO lightcurves
    is still meager, with few of sufficient quality for analysis (see
    Table~\ref{TabKBO}). This is due to the fact that most KBOs are
    faint objects, with apparent red magnitude of $m_{\rm
    R}\!\!\sim\!\!23$ \citep{trujillo}, rendering it very difficult to
    detect small amplitude changes in their brightness. One of the few
    high quality lightcurves is that of (20000) Varuna, which shows an
    amplitude of $\Delta m=0.42\pm0.02$~mag and a period of $P_{\rm
    rot}=6.3442\pm0.0002$~hrs \citep{jewitt4}. Only recently have
    surveys started to yield significant numbers of KBOs bright enough
    for detailed studies \citep{jewitt2}.

    Another difficulty associated with the measurement of the
    amplitude of a lightcurve is the one of determining the period of
    the variation. If no periodicity is apparent in the data, any
    small variations in the brightness of an object must be due to
    noise. Furthermore, a precise measurement of the amplitude of the
    lightcurve requires a complete coverage of the rotational
    phase. Therefore, any conclusion based on amplitudes of
    lightcurves must assume that their periods have been determined
    and confirmed by well sampled phase plots of the data.
    
    However, not all of the observed KBOs show detectable brightness
    variations (the so-called ``flat'' lightcurves). The simplest
    explanations for this could be due to (1) the object is
    axisymmetric (the two axes perpendicular to the spin vector are
    equal), or (2) its rotation axis is nearly coincident with the
    line of sight (see Fig.~\ref{FigAA}). In other words, the
    undetectable variations are either a consequence of the KBO's
    shape, or of the observational geometry. By studying the relative
    probabilities of these two causes, and relating them to the
    observed fraction of "flat" lightcurves, we might expect to
    improve our knowledge of the intrinsic shape distribution of
    KBOs. In this paper we address the following question: Can we
    learn something about the shape distribution of KBOs from the
    fraction of ``flat'' lightcurves?


\begin{deluxetable}{lccr@{.}lr@{.}lrccl}
  \tablecolumns{12}
  \tabletypesize{\small}
  \tablecaption{KBOs with measured lightcurves. \label{TabKBO}}
  \tablehead{
    \colhead{Name} & \colhead{Class\tablenotemark{a}} &
    \colhead{$H$\tablenotemark{b}} & \multicolumn{2}{c}{$\Delta
      m$\tablenotemark{c}} & \multicolumn{2}{c}{$P$\tablenotemark{d}} & 
    \colhead{Source\tablenotemark{e}}\\
    &  & \colhead{[mag]} & \multicolumn{2}{c}{[mag]} &
    \multicolumn{2}{c}{[hrs]} & 
  }
  \startdata
  1993$\,$SC        & C & 6.9 & $<$0&04 & \multicolumn{2}{c}{ } & 
  RT99\\ 
  1994$\,$TB        & P & 7.1 &    0&3  & 6&5                   & 
  RT99 \\ 
  1996$\,$TL$_{66}$ & S & 5.4 & $<$0&06 & \multicolumn{2}{c}{ } &
  RT99 \\ 
  1996$\,$TP$_{66}$ & P & 6.8 & $<$0&12 & \multicolumn{2}{c}{ } &
  RT99\\ 
  1994$\,$VK$_{8}$  & C & 7.0 &    0&42 & 9&0                   & 
  RT99 \\
  1996$\,$TO$_{66}$ & C & 4.5 &    0&1  & 6&25                  & 
  Ha00 \\
  Varuna            & C & 3.7 &    0&42 & 6&34                  & 
  JS02 \\ 
  1995$\,$QY$_{9}$  & P & 7.5 &    0&6  & 7&0                   & 
  RT99\\ 
  1996$\,$RQ$_{20}$ & C & 7.0 &\multicolumn{2}{c}{-}&\multicolumn{2}{c}{ }&
  RT99\\ 
  1996$\,$TS$_{66}$ & C & 6.4 & $<$0&16 & \multicolumn{2}{c}{ } & 
  RT99\\ 
  1996$\,$TQ$_{66}$ & C & 7.0 & $<$0&22 & \multicolumn{2}{c}{ } &
  RT99\\ 
  1997$\,$CS$_{29}$ & C & 5.2 & $<$0&2 & \multicolumn{2}{c}{ } &
  RT99\\ 
  1999$\,$TD$_{10}$ & S & 8.8 & 0&68 & 5&8 &
  Co00\\
  \enddata
  \tablenotetext{a}{\footnotesize dynamical class (C - classical
    KBO, P - plutino, S - scattered KBO)}
  \tablenotetext{b}{\footnotesize absolute magnitude}
  \tablenotetext{c}{\footnotesize lightcurve amplitude}
  \tablenotetext{d}{\footnotesize rotational period}
  \tablenotetext{e}{\footnotesize JS02 \citep{jewitt4}, Ha00
    \citep{hainaut}, Co00 \citep{consolmagno}, RT99
    \citep{romanishin}}
\end{deluxetable}

    \begin{deluxetable}{rl}
      \tabletypesize{\small}
      \tablecolumns{2}
      \tablecaption{Used symbols and notation. \label{TabNot}}
      \tablehead{
	\colhead{Symbol} & \colhead{Description}
      }
      \startdata
$a\!\geq\!b\!\geq\!c$ & axes of ellipsoidal KBO\\
      $\bar{a}\!\geq\!\bar{b}\!\geq\!\bar{c}$ & 
      normalized axes of KBO $\;(\bar{b}=1)$\\
      $\theta$     & aspect angle \\
      $\Delta m_{\rm min}$ & minimum detectable lightcurve amplitude \\
      $\theta_{\rm min}$ &aspect angle at which $\Delta m=\Delta m_{\rm min}$\\
      $K$          & $10^{0.8\Delta m_{\rm min}}$      
      \enddata
    \end{deluxetable}


\section{Definitions and Assumptions}

    The observed brightness variations in KBO lightcurves can be due to:
    \begin{list}{$\cdot$}{\itemsep 0cm}
    \item eclipsing binary KBOs
    \item surface albedo variations
    \item irregular shape
    \end{list}
    In general the brightness variations will arise from some
    combination of these three factors, but the preponderance of each
    effect among KBOs is still not known. In the following
    calculations we exclude the first two factors and assume that
    shape is the sole origin of KBO brightness variations. We further
    assume that KBO shapes can be approximated by triaxial ellipsoids,
    and thus expect a typical KBO lightcurve to show a set of 2 maxima
    and 2 minima for each full rotation (see Fig.~\ref{FigLC}).
    Table~\ref{TabNot} summarizes the used symbols and notations. The
    listed quantities are defined in the text. 

    The detailed assumptions of our model are as follows:

    \begin{figure}
      \epsscale{1.0}
      \plotone{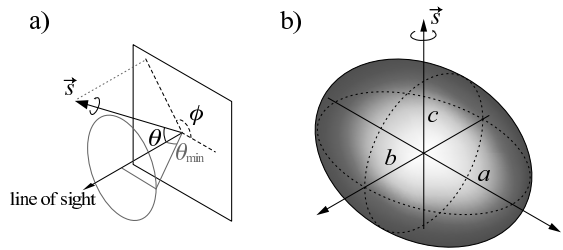}
      \caption{ {\bf a)} A spherical coordinate system is used to
	represent the observing geometry. The line of sight (oriented
	from the object to the observer) is the polar axis and the
	azimuthal axis is arbitrary in the plane orthogonal to the
	polar axis. $\theta$ and $\phi$ are the spherical angular
	coordinates of the spin axis $\vec{s}$. In this coordinate
	system the aspect angle is given by $\theta$. The
	\mbox{``non-detectability''} cone, with semi-vertical angle
	$\theta_{\rm min}$, is represented in grey. If the spin axis
	lies within this cone the brightness variations due to
	changing cross-section will be smaller than photometric
	errors, rendering it impossible to detect brightness
	variations. {\bf b)} The picture represents an ellipsoidal KBO
	with axes ${a\geq b\geq c}$.
	\label{FigEeC}}
    \end{figure}

    \begin{enumerate}
    \item {\em The KBO shape is a triaxial ellipsoid.} This is the
      shape assumed by a rotating body in hydrostatic equilibrium
      \citep{chandra}. There are reasons to believe that KBOs might
      have a ``rubble pile'' structure \citep{farinella}, justifying
      the approximation even further.

    \item {\em The albedo is constant over surface.} Although albedo
      variegation can in principle explain any given lightcurve
      \citep{russell}, the large scale brightness variations are
      generally attributed to the object's irregular shape
      \citep{burns79}.

    \item {\em All axis orientations are equally probable.} Given that
      we have no knowledge of preferred spin vector orientation, this
      is the most reasonable {\it a priori} assumption.

    \item {\em The KBO is in a state of simple rotation around the
      shortest axis (the axis of maximum moment of inertia)}. This is
      likely since the damping timescale of a complex rotation (e.g.,
      precession), $\sim10^{3}\,$yr, \citep{burns}, \citep{harris94}
      is smaller than the estimated time between collisions
      ($10^{7}$--$10^{11}\,$yr) that would re-excite such a rotational
      state \citep{stern}, \citep{davis}.

    \item {\em The KBO is observed at zero phase angle} ($\alpha=0$)
      It has been shown from asteroid data that lightcurve amplitudes
      seem to increase linearly with phase angle,
      \begin{eqnarray*}
	A(\theta,0)=A(\theta,\alpha)/(1+m\alpha)\;,
      \end{eqnarray*}
      where $\theta$ is the aspect angle, $\alpha$ is the phase angle
      and $m$ is a coefficient which depends on surface
      composition. The aspect angle is defined as the angle between
      the line of sight and the spin axis of the KBO (see
      Fig.~\ref{FigEeC}a), and the phase angle is the Sun-object-Earth
      angle. The mean values of $m$ found for different asteroid
      classes are ${m({\rm S})=0.030}$, ${m({\rm C})=0.015}$, ${m({\rm
      M})=0.013}$, where S, C, and M are asteroid classes
      \citep{micha}. Since KBO are distant objects the phase angle
      will always be small. Even allowing $m$ to be one order of
      magnitude higher than that of asteroids the increase in the
      lightcurve amplitude will not exceed 1\%.

    \item {\em The brightness of the KBO is proportional to its
      cross-section area (geometric scattering law)}. This is a good
      approximation for KBOs because (1) most KBOs are too small to
      hold an atmosphere, and (2) the fact that they are observed at
      very small phase angles reduces the influence of scattering on
      the lightcurve amplitude \citep{magnusson89}.
    \end{enumerate}

    The KBOs will be represented by triaxial ellipsoids of axes ${a
    \geq b \geq c}$ rotating around the short axis $c$ (see
    Fig.~\ref{FigEeC}b). In order to avoid any scaling factors we
    normalize all axes by $b$, thus obtaining a new set of parameters
    $\bar{a}$, $\bar{b}$ and $\bar{c}$ given by
    \begin{eqnarray}
      \bar{a}=a/b\;,\;\;\; \bar{b}=1\;, \;\;\; \bar{c}=c/b\;.
    \end{eqnarray}
    As defined, $\bar{a}$ and $\bar{c}$ can assume values
    ${1\leq\bar{a}<\infty}$ and ${0<\bar{c}\leq 1}$. Note that the
    parameters $\bar{a}$ and $\bar{c}$ are dimensionless.

    The orientation of the spin axis of the KBO relative to the line
    of sight will be defined in spherical coordinates $(\theta,\phi)$,
    with the line of sight (oriented from the object to the observer)
    being the $z$-axis, or polar axis, and the angle $\theta$ being
    the polar angle (see Fig.~\ref{FigEeC}a). The solution is
    independent of the azimuthal angle $\phi$, which would be measured
    in the plane perpendicular to the line of sight, between an
    arbitrary direction and the projection of the spin axis on the
    same plane. The observation geometry is parameterized by the
    aspect angle, which in this coordinate system corresponds to
    $\theta$.
    
    As the object rotates, its cross-section area $S$ will vary
    periodically between $S_{\rm max}$ and $S_{\rm min}$ (see
    Fig.~\ref{FigLC}). These areas are simply a function of $a$, $b$,
    $c$ and the aspect angle $\theta$. Given the assumption of
    geometric scattering, the ratio between maximum and minimum flux
    of reflected sunlight will be equal to the ratio between $S_{\rm
    max}$ and $S_{\rm min}$. The lightcurve amplitude can then be
    calculated from the quantities $\bar{a}$, $\bar{c}$ and $\theta$
    and is given by
    \begin{eqnarray}
      \Delta m=2.5\log \left(
      \frac{\bar{a}^{2}\cos^{2}\theta+\bar{a}^{2}\bar{c}^{2}\sin^{2}\theta}
      {\bar{a}^{2}\cos^{2}\theta+\bar{c}^{2}\sin^{2}\theta}
      \right)^{1/2}\;.
      \label{EqnDeltaMag}
    \end{eqnarray}


\section{``Flat'' Lightcurves}

    It is clear from Eq.~(\ref{EqnDeltaMag}) that under certain
    conditions, $\Delta m$ will be zero, i.e., the KBO will exhibit a
    flat lightcurve. These special conditions involve the shape of the
    object and the observation geometry, and are described
    quantitatively below. Taking into account photometric error bars
    will bring this ``flatness'' threshold to a finite value, $\Delta
    m_{\rm min}$, a minimum detectable amplitude below which
    brightness variation cannot be ascertained.

    The two factors that influence the amplitude of a KBO lightcurve are:

    {\bf 1. Sphericity} For a given ellipsoidal KBO of axes ratios
    $\bar{a}$ and $\bar{c}$ the lightcurve amplitude will be largest
    when $\theta=\pi/2$ and smallest when $\theta=0$ or $\pi$. At
    $\theta=\pi/2$, Eq.~(\ref{EqnDeltaMag}) becomes
    \begin{eqnarray}
      \Delta m=2.5\log\bar{a}\;.
      \label{EqnDeltaMagPerp}
    \end{eqnarray}
    Even at $\theta=\pi/2$, having a minimum detectable amplitude,
    $\Delta m_{\rm min}$, puts constraints on $\bar{a}$ since if
    $\bar{a}$ is too small, the lightcurve amplitude will not be
    detected. This constraint is thus
    \begin{eqnarray}
      \bar{a}<10^{0.4\Delta m_{\rm min}} \Rightarrow \mbox{``flat''
      lightcurve}\;.
      \label{EqnCondFLC}
    \end{eqnarray}

    \begin{figure}
      \epsscale{1.0} \plotone{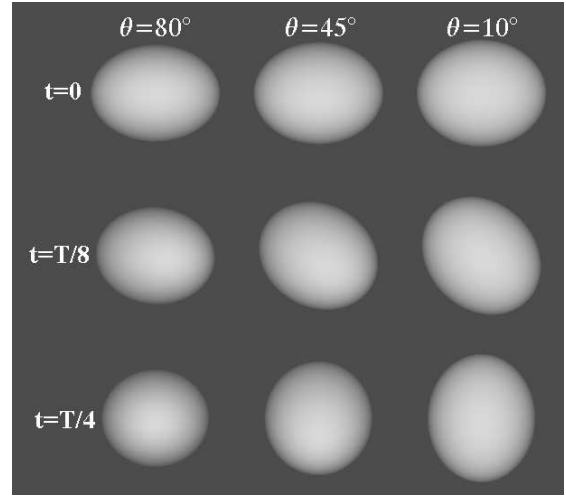}
      \caption{Illustration of a rotating ellipsoid at different
	aspect angles. A quarter of a full rotation is
	represented. Rotational phase of ellipsoid increasing from top
	to bottom and $\theta$ decreasing from left to right. T is the
	period of rotation. Axes ratios are $\bar{a}=1.2$ and
	$\bar{c}=0.9$. \label{FigAA}}
    \end{figure}

    {\bf 2. Observation geometry} If the rotation axis is nearly
    aligned with the line of sight, i.e., if the aspect angle is
    sufficiently small, the object's projected cross-section will
    hardly change with rotation, yielding no detectable brightness
    variations (see Fig.~\ref{FigAA}). The finite accuracy of the
    photometry defines a minimum aspect angle, $\theta_{\rm min}$,
    within which the lightcurve will appear flat within the
    uncertainties. This angle rotated around the line of sight
    generates the \mbox{``non-detectability cone''} (see
    Fig.~\ref{FigEeC}a), with the solid angle
    \begin{eqnarray}
      \Omega(\theta_{\rm min})=\int^{2\pi}_0\!\int^{\theta_{\rm
      min}}_0 \sin\theta\,\,\mathrm{d}\theta \,\mathrm{d}\phi\;.
    \end{eqnarray}
    Any aspect angle $\theta$ which satisfies $\theta<\theta_{\rm
    min}$ falls within the ``non-detectability cone'' and results in a
    non-detectable lightcurve amplitude. Therefore, the probability
    that the lightcurve will be flat due to observing geometry is
    \begin{subequations}
    \begin{eqnarray}
        p_{\bar{a},\bar{c}}(\mbox{non-detection}) & = &
	\frac{2\times\Omega(\theta_{\rm min})}{4\pi} \nonumber\\
        & = & 1-\cos\theta_{\rm min} \\
        p_{\bar{a},\bar{c}}(\mbox{detection}) & = &
        \cos\theta_{\rm min}.
    \end{eqnarray}
    \end{subequations}

    The factor of 2 accounts for the fact that the axis might be
    pointing towards or away from the observer and still give rise to
    the same observations, and the $4\pi$ in the denominator
    represents all possible axis orientations.

\begin{figure}
      \epsscale{1.15}
      \plotone{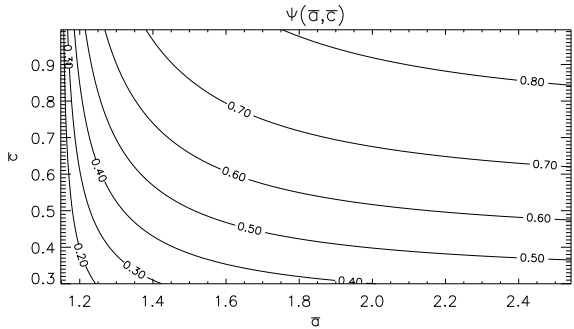}
      \caption{The function $\Psi(\bar{a},\bar{c})$
	(Eq.~(\ref{EqnPsi})). This plot assumes photometric errors
	${\Delta m_{\rm min}=0.15\,}$mag. The detection probability is
	zero when ${\bar{a}<10^{0.4\Delta m_{\rm min}}\approx 1.15}$.
	\label{FigPsi}}
    \end{figure}

    From Eq.~(\ref{EqnDeltaMag}) we can write $\cos \theta_{\rm min}$
    as a function of $\bar{a}$ and $\bar{c}$,
    \begin{eqnarray}
        \cos\theta_{\rm min} = \Psi(\bar{a},\bar{c}) =
	\sqrt{\frac{\bar{c}^{2}(\bar{a}^{2}-K)}
	{\bar{c}^{2}(\bar{a}^{2}-K)+\bar{a}^{2}(K-1)}},
        \label{EqnPsi}
    \end{eqnarray}
    where $K=10^{0.8\Delta m_{\rm min}}$. The function
    $\Psi(\bar{a},\bar{c})$, represented in Fig.~\ref{FigPsi}, is the
    probability of detecting brightness variation from a given
    ellipsoid of axes ratios $(\bar{a},\bar{c})$. It is a geometry
    weighting function. For $\bar{a}$ in $[1,\sqrt{K}]$ we have
    ${\Psi(\bar{a},\bar{c})=0}$ by definition, since in this case the
    KBO satisfies Eq.~(\ref{EqnCondFLC}) and its lightcurve amplitude
    will not be detected irrespective of the aspect angle. It is clear
    from Fig.~\ref{FigPsi} that it is more likely to detect brightness
    variation from an elongated body.

\section{Detectability of Lightcurves}

    In order to generate a ``non-flat'' lightcurve, the KBO has to
    satisfy both the shape and observation geometry conditions.
    Mathematically this means that the probability of detecting
    brightness variation from a KBO is a function of the probabilities
    of the KBO satisfying both the sphericity and observing geometry
    conditions.

    We will assume that it is possible to represent the shape
    distribution of KBOs by two independent probability density
    functions, $f(\bar{a})$ and $g(\bar{c})$,
    defined as
    \begin{subequations}
    \begin{eqnarray}
      \label{EqnSDDef}
      p(\bar{a}_1\!\leq\!\bar{a}\!\leq\!\bar{a}_2) =
      \int^{\bar{a}_2}_{\bar{a}_1} f(\bar{a})\,\mathrm{d}\bar{a} &,&
      \int^\infty_1 f(\bar{a})\,\mathrm{d}\bar{a}=1, \\
      p(\bar{c}_1\!\leq\!\bar{c}\!\leq\!\bar{c}_2) =
      \int^{\bar{c}_2}_{\bar{c}_1} g(\bar{c})\,\mathrm{d}\bar{c} &,&
      \int^1_0 g(\bar{c})\,\mathrm{d}\bar{c}=1,
    \end{eqnarray}
    \end{subequations}
    where the integrals on the left represent the fraction of KBOs in
    the given ranges of axes ratios.  This allows us to write the
    following expression for $p(\Delta m>\Delta m_{\rm min})$, where
    both the shape and observation geometry constraints are taken into
    account,
    \begin{eqnarray}
      \lefteqn{p(\Delta m>\Delta m_{\rm min})=} \nonumber \\
      & & \int^1_0 \int^{\infty}_1\Psi(\bar{a},\bar{c})
      f(\bar{a})g(\bar{c}) \,\mathrm{d}\bar{a}\,\mathrm{d}\bar{c}.
      \label{EqnProb}
    \end{eqnarray}
    The right hand side of this equation represents the probability of
    observing a given KBO with axes ratios between $(\bar{a},\bar{c})$
    and $(\bar{a}+\mathrm{d}\bar{a},\bar{c}+\mathrm{d}\bar{c})$, at a
    large enough aspect angle, integrated for all possible axes
    ratios. This is also the probability of detecting brightness
    variation for an observed KBO.
    
    The lower limit of integration for $\bar{a}$ in
    Eq.~(\ref{EqnProb}) can be replaced by $\sqrt{K}$, with $K$
    defined as in Eq.~(\ref{EqnPsi}), since $\Psi(\bar{a},\bar{c})$
    is zero for $\bar{a}$ in $[1,\sqrt{K}]$. In fact, this is how the
    sphericity constraint is taken into account.

    Provided that we know the value of $p(\Delta m>\Delta m_{\rm
    min})$ Eq.~(\ref{EqnProb}) can test candidate distributions
    $f(\bar{a})$ and $g(\bar{c})$ for the shape distribution of
    KBOs. The best estimate for $p(\Delta m>\Delta m_{\rm min})$ is
    given by the ratio of ``non-flat'' lightcurves ($N_{\rm D}$) to
    the total number of measured lightcurves ($N$), i.e.,
    \begin{eqnarray}
      p(\Delta m>\Delta m_{\rm min})\approx\frac{N_{\rm D}}{N}.
      \label{EqnNFLC}
    \end{eqnarray}
    Because $N$ is not the total number of KBOs there will be an error
    associated with this estimate. Since we do not know the
    distributions $f(\bar{a})$ and $g(\bar{c})$ we will assume that
    the outcome of an observation can be described by a binomial
    distribution of probability ${p(\Delta m>\Delta m_{\rm
    min})}$. This is a good approximation given that $N$ is very small
    compared with the total number of KBOs. Strictly speaking, the
    hypergeometric distribution should be used since we will not
    unintentionally observe the same object more than once (sampling
    without replacement). However, since the total number of KBOs
    (which is not known with certainty) is much larger than any sample
    of lightcurves, any effects of repeated sampling will be
    negligible, thereby justifying the binomial approximation.  This
    simplification allows us to calculate the upper ($p_+$) and lower
    ($p_-$) limits for $p(\Delta m>\Delta m_{\rm min})$ at any given
    confidence level, $C$. These values, known as the Clopper--Pearson
    confidence limits, can be found solving the following equations by
    trial and error \citep{barlow},
    \begin{subequations}
    \begin{eqnarray}
        \sum_{r=N_{\rm D}+1}^{N} P\big(r;p_+(\Delta m\!>\!\Delta
        m_{\rm min}),N\big) =\frac{C+1}{2} \\ 	
	\sum_{r=0}^{N_{\rm D}-1} P\big(r;p_-(\Delta m\!>\!\Delta
        m_{\rm min}),N\big) = \frac{C+1}{2},
    \end{eqnarray}
    \end{subequations}
    \noindent (see Table~\ref{TabNot} for notation) where $C$ is the desired
    confidence level and $P(r;p,N)$ is the binomial probability of
    detecting $r$ lightcurves out of $N$ observations, each lightcurve
    having a detection probability $p$. Using the values in
    Table~\ref{TabKBO} and $\Delta m_{\rm min}=0.15\,$mag we
    have $N_{\rm D}=5$ and $N=13$ which yields ${p(\Delta m>\Delta
    m_{\rm min})=0.38^{+0.18}_{-0.15}}$ at a $C=0.68$ $(1\sigma)$
    confidence level. At $C=0.997$ $(3\sigma)$ we have ${p(\Delta
    m>\Delta m_{\rm min})=0.38^{+0.41}_{-0.31}}$. The value of
    ${p(\Delta m>\Delta m_{\rm min})}$ could be smaller since some of
    the flat lightcurves might not have been published.

\begin{figure}
      \epsscale{1.15}
      \plotone{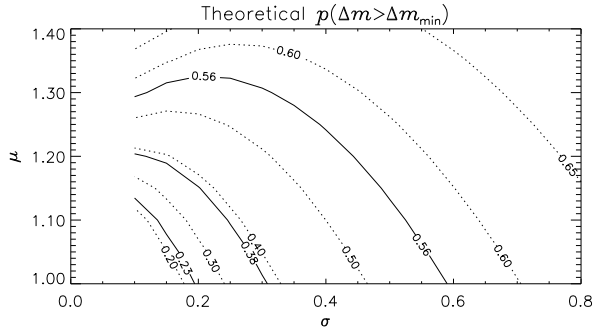}
      \caption{Contour plot of the theoretical probabilities of
	detecting brightness variation in KBOs (assuming ${\Delta
	m_{\rm min}=0.15\,}$mag), drawn from gaussian shape
	distributions parameterized by $\mu$ and $\sigma$
	(respectively the mean and spread of the distributions). The
	solid lines represent the observed ratio of ``non-flat''
	lightcurves (at 0.38) and 0.68 confidence limits (at 0.23 and
	0.56 respectively). \label{FigThp}}
    \end{figure}

    Note that for moderately elongated ellipsoids (small $\bar{a}$)
    the function $\Psi(\bar{a},\bar{c})$ is almost insensitive to the
    parameter $\bar{c}$ (see Fig.~\ref{FigPsi}),
    in which case the axisymmetric approximation with respect to
    $\bar{a}$ can be made yielding $\bar{c}\approx
    1$. Eq.~(\ref{EqnProb}) then has only one unknown parameter,
    $f(\bar{a})$.
    \begin{eqnarray}
        p(\Delta m > \Delta m_{\rm min}) & \approx &
        \int^{\bar{a}_{\rm max}}_{\sqrt{K}} \Psi(\bar{a},1)
        f(\bar{a})\,\mathrm{d}\bar{a} \nonumber \\
	& \approx & 0.38^{+0.41}_{-0.31}.
        \label{EqnProb2}
    \end{eqnarray}
    If we assume the function $f(\bar{a})$ to be gaussian, we can use
    Eq.~(\ref{EqnProb2}) to determine its mean $\mu$ and standard
    deviation $\sigma$, after proper normalization to satisfy
    Eq.~(\ref{EqnSDDef}).  The result is represented in
    Fig.~\ref{FigThp}, where we show all possible pairs of
    ($\mu$,$\sigma$) that would satisfy a given $p(\Delta m>\Delta
    m_{\rm min})$.  For example, the line labeled "0.38" identifies
    all possible pairs of ($\mu$,$\sigma$) that give rise to $p(\Delta
    m>\Delta m_{\rm min}) = 0.38$, the line labeled "0.56" all
    possible pairs of ($\mu$,$\sigma$) that give rise to $p(\Delta
    m>\Delta m_{\rm min} ) = 0.56$, etc.

    Clearly, with the present number of lightcurves the uncertainties
    are too large to draw any relevant conclusions on the shape
    distribution of KBOs. With a larger dataset, this formulation will
    allow us to compare the distribution of KBO shapes with that of
    the main belt asteroids. The latter has been shown to resemble, to
    some extent, that of fragments of high-velocity impacts
    \citep{catullo}. It deviates at large asteroid sizes that have
    presumably relaxed to equilibrium figures.  A comparison of
    $f(\bar{a})$ with asteroidal shapes should tell us, at the very
    least, whether KBO shapes are collisionally derived, as opposed to
    being accretional products.

    The usefulness of this method is that, with more data, it would
    allow us to derive such quantitative parameters as the mean and
    standard deviation of the KBO shape distribution, if we assume
    {\it a priori} some intrinsic form for this distribution.  The
    method's strength is that it relies solely on the detectability of
    lightcurve amplitudes, which is more robust than other lightcurve
    parameters.

    This paper focuses on the influence of the observation geometry
    and KBO shapes in the results of lightcurve measurements. In which
    direction would our conclusions change with the inclusion of
    albedo variegation and/or binary KBOs?

    Non-uniform albedo would cause nearly spherical KBOs to generate
    detectable brightness variations, depending on the coordinates of
    the albedo patches on the KBO's surface. This means that our
    method would overestimate the number of elongated objects by
    attributing all brightness fluctuations to asphericity.

    Binary KBOs would influence the results in different ways
    depending on the orientation of the binary system's orbital plane,
    on the size ratio of the components, and on the individual shapes
    and spin axis orientations of the primary and secondary. For
    example, an elongated KBO observed equator-on would have its
    lightcurve flattened by a nearly spherical moon orbiting in the
    plane of the sky, whereas two spherical KBOs orbiting each other
    would generate a lightcurve if the binary would be observed
    edge-on.
    
    These effects are not straightforward to quantify analytically and
    might require a different approach. We intend to incorporate them
    in a future study. Also, with a larger sample of lightcurves it
    would be useful to apply this model to subgroups of KBOs based on
    dynamics, size, etc.

\section{Conclusions}

    We derived an expression for the probability of detecting
    brightness variations from an ellipsoidal KBO, as a function of
    its shape and minimum detectable amplitude. This expression takes
    into account the probability that a ``flat'' lightcurve is caused
    by observing geometry.

    Our model can yield such quantitative parameters as the mean and
    standard deviation of the KBO shape distribution, if we assume
    {\it a priori} an intrinsic form for this distribution. It
    concerns solely the statistical probability of detecting
    brightness variation from objects drawn from these distributions,
    given a minimum detectable lightcurve amplitude. The method relies
    on the assumption that albedo variegation and eclipsing binaries
    play a secondary role in the detection of KBO lightcurves. The
    effect of disregarding albedo variegation in our model is that we
    might overestimate the fraction of elongated objects. Binaries in
    turn could influence the result in both directions depending on
    the geometry of the problem, and on the physical properties of the
    constituents. We intend to incorporate these effects in a future,
    more detailed study.

\acknowledgments

We are grateful to Garrelt Mellema, Glenn van de Ven, and Prof. John
Rice for helpful discussion.


\end{document}